\documentstyle[amsfonts,11pt]{article}

\def\bbt{\bibitem}
\def\be{\begin{equation}}
\def\en{\end{equation}}
\def\ber{\begin{eqnarray}}
\def\enr{\end{eqnarray}}
\def\nmb{ \nonumber\\}
\def\d{\partial}
\def\rbr{\rbrack}
\def\lbr{\lbrack}

\def\ov{\over }
\def\tld{\tilde}

\def\Sgm{\Sigma}
\def\al{\alpha}
\def\bt{\beta}
\def\gm{\gamma}
\def\Gm{\Gamma}
\def\im{\imath}

\def\et{\eta}
\def\tt{\theta}

\def\dlt{\delta}
\def\Dl{\Delta}
\def\kp{\kappa}

\begin{document}
\rightline{Landau Tmp/02/01.}
\rightline{February 2001}
\vskip 2 true cm
\centerline{\bf BRST CONSTRUCTION OF $D$-BRANES IN $SU(2)$ WZW MODEL.}
\vskip 1.5 true cm
\centerline{\bf S. E. Parkhomenko}
\centerline{Landau Institute for Theoretical Physics}
\centerline{142432 Chernogolovka,Russia}
\vskip 0.5 true cm
\centerline{spark@itp.ac.ru}
\vskip 1 true cm
\centerline{\bf Abstract}
\vskip 0.5 true cm

\centerline{BRST construction of $D$-branes in $SU(2)$ WZW model
is proposed.}
\vskip 10pt

"{\it PACS: 11.25Hf; 11.25 Pm.}"


\smallskip
\vskip 10pt
\centerline{\bf1. Introduction}
\vskip 10pt

 The role of $D$-branes ~\cite{Pol} in the description of certain
non perturbative degrees of freedom of strings is by now well
established and the study of their dynamics has lead to many
new insights into string  and $M$-theory ~\cite{Wit}.
Much of this study was done in the large volume regime where
geometric techniques provide reliable information. The extrapolation
into the stringy regime usually requires boundary conformal field
theory (CFT) methods. In this approach $D$-brane configurations are
given by conformally invariant boundary states or boundary conditions.
However the geometric properties of these configurations are well
understood only for the case of flat and toric backgrounds where
the CFT on the world sheet is a theory of free fields. Due to this
reason many calculations concerning the scattering and emission of
closed strings from $D$-branes can be given exactly
(see for example ~\cite{FrPSLR}, ~\cite{HINS}). In particular,
geometric features of the $D$-brane solutions of supergravity
in flat 10-dimensional space-time were recovered from the boundary
states in ~\cite{DiVe}.

 The class of rational CFT's gives the examples of curved
string backgrounds where the construction of the boundary states
leaving the whole chiral symmetry algebra or its subalgebra unbroken can
be given in principle and the interaction of these states
with closed strings can be calculated exactly. However the
extraction of
geometric features of boundary states is a problem as against
the flat string background, because the boundary states
are given in purely algebraic dates of the rational CFT of interest.
One of the most important examples of this situation is given
by Gepner models ~\cite{Gep}, where the boundary state approach
developed in ~\cite{ReSch}, ~\cite{Qint}, ~\cite{DCY}
has been used to get some of
the geometric features of $D$-branes at small volume of
the Calabi-Yau manifold.

 WZW models on the compact groups is a subclass of rational
CFT's where the chiral symmetry algebra is given by a Kac-Moody
algebra and the space of states is given by the direct sum of
integrable representations of the algebra ~\cite{Kac}.
$D$-branes in these
models have been intensively studied in the last few years
~\cite{KlS}-~\cite{RR}. In particular,
geometry of the boundary states which left unbroken
Kac-Moody algebra of symmetries has been investigated
in ~\cite{FFFS}.

 There is well known free field realization of WZW models
~\cite{BFeld}-~\cite{GMOM}
which allows to describe the irreducible representations
of the Kac-Moody algebras, vertex operators and calculate
the correlation functions. It is natural to ask
whether we can extend the free field realization for boundary WZW
models. To answer this question one needs to find first
a free field construction of the boundary states.

 This problem has been treated recently in ~\cite{IW}.
In this work
the known expression of a character of the irreducible
representation of $su(2)$ Kac-Moody algebra via
an alternating sum of the characters of Fock modules ~\cite{BFeld}
has been used
to represent Ishibashi state of the irreducible $\hat{su}(2)$
representation as a superposition of
Isibashi states of the Fock modules.
But this is certainly not the end of the construction
because one needs to check that the boundary states
constructed this way do not emit non-physical closed string states
which are present originally in the free field realization.
Similar to the bulk situation,
the condition that non-physical states are not radiated by
the boundary state is equivalent to BRST
invariance of the state, with respect to the sum of BRST charges
of Felder's resolution in
the left- and right-moving sectors of the model.
This condition which is of crucial importance in the free field
construction of boundary states was not taken into account
in ~\cite{IW}.

 In this note we study this problem for $SU(2)$ WZW model.
In Section 2 we follow closely to the Section 3 of ~\cite{IW}
and construct $\hat{su}(2)$ invariant Ishibashi
states in the tensor product of Wakimoto modules from left-
and right-moving sectors.
Section 3 is the main part of the present note. In this section we
construct Ishibashi state for each irreducible integrable
$\hat{su}(2)$ representation using the superpositions
of Ishibashi states of Wakimoto modules from Felder's resolution.
We find the coefficients
of the superposition imposing the BRST invariance condition
which is similar to that of the bulk theory.
Then we show that Ishibashi state constructed this way is not
BRST exact and hence, represents a nontrivial homology class.
At the end of Section 3 the boundary states in $SU(2)$ WZW model
are constructed using the solution found by Cardy ~\cite{C}.
In Section 4 we consider briefly free field realization
of the twisted boundary states and construct twisted
BRST invariant boundary state for the case
of Weyl reflection automorphism. Section 5 is devoted to
discussions.

\smallskip
\vskip 10pt
\centerline{\bf2. Ishibashi states in $\hat {sl}(2)$ Wakimoto modules}
\vskip 10pt

 The standard $\hat {sl}(2)$-Wakimoto
representation ~\cite{Wak} in the left-moving sector is given by
the set of free scalar fields $a(z)$, $a^{+}(z)$, $\phi(z)$:
\ber
a^{+}(z_{1})a(z_{2})=-z_{12}^{-1}+reg., \nmb
\phi(z_{1})\phi(z_{2})=-\ln z_{12}+reg.,
\label{1.1}
\enr
\ber
F=-a^{+}, \nmb
H=-2aa^{+}+\im \sqrt{2}\nu \d \phi, \nmb
E=a^{2}a^{+}+k\d a-\im \sqrt{2}\nu a\d \phi,
\label{1.2d}
\enr
\ber
T=T_{\phi}+T_{a}, \nmb
T_{\phi}=-{1\ov 2}(\d \phi)^{2}+
\im {1\ov \sqrt{2}\nu}\d^{2}\phi, \ T_{a}=-a^{+}\d a
\label{1.Td}
\enr

The parameter $\nu$ is related with the level $k$ of the
$\hat {sl}(2)$, $\nu^{2}=k+2$.
Similar to (\ref{1.1}), (\ref{1.2d}), Wakimoto representation in the
right-moving sector is given
by the set of analogous formulas with the opposit sign in front of
$\bar{\d} \bar{\phi}$ ~\cite{IW}.
The OPE's (\ref{1.1}) give the commutation relations for the
modes of the left-moving fields:
\ber
\lbr a(n),a^{+}(m)\rbr=\dlt_{n+m,0}, \nmb
\lbr \phi(n),\phi(m)\rbr=-n\dlt_{n+m,0}.
\label{1.4}
\enr
(the similar relations are valid for the modes in the right-moving
sector).
Let us denote by $|P,l>$ the vacuum state of the fields $a,a^{+},\phi$
which fulfills the relations
\ber
a(n)|P,l>=0, \ n>P, \nmb
a^{+}(n)|P,l>=0, \ n>-P-1,
\label{1.5}
\enr
\ber
\phi(n)|P,l>=0, \ n>0 \nmb
\phi(0)|P,l>=\im {l \ov \sqrt{2}\nu}|P,l>
\label{1.6}
\enr
( P is a picture number ~\cite{FMS}).
This state has the following properties:
\ber
F(n)|P,l>=0,\ n>-1-P, \nmb
E(n)|P,l>=0,\ n>P, \nmb
H(n)|P,l>=0, \ n>0, \ H(0)|P,l>=(2P-l)|P,l>, \nmb
L(n)|P,l>=0, \ n>0, \nmb 
L(0)|P,l>=(-{P(P-1)\ov 2}+{l(l+2)\ov 4\nu^{2}})|P,l>.
\label{1.6KMd}
\enr

 The state in the right-moving sector $|\bar{P},\bar{l}>$
fulfills the relations
\ber
\bar{a}(n)|\bar{P},\bar{l}>=0, \ n>\bar{P}, \nmb
\bar{a}^{+}(n)|\bar{P},\bar{l}>=0, \ n>-\bar{P}-1,
\label{1.5bar}
\enr
\ber
\bar{\phi}(n)|\bar{P},\bar{l}>=0, \ n>0 \nmb
\bar{\phi}(0)|\bar{P},\bar{l}>=
\im {\bar{l} \ov \sqrt{2}\nu}|\bar{P},\bar{l}>.
\label{1.6bar}
\enr
This state has the following properties:
\ber
\bar{F}(n)|\bar{P},\bar{l}>=0, \ n>-\bar{P}-1, \nmb
\bar{E}(n)|\bar{P},\bar{l}>=0, \ n>\bar{P}, \nmb
\bar{H}(n)|\bar{P},\bar{l}>=0, \ n>0, \
\bar{H}(0)|P,l>=(2\bar{P}+\bar{l})|\bar{P},\bar{l}>, \nmb
\bar{E}(n)|\bar{P},\bar{l}>=0, \ n>\bar{P}, \nmb
\bar{L}(n)|\bar{P},\bar{l}>=0, \ n>0, \
\bar{L}(0)|\bar{P},\bar{l}>=(-{\bar{P}(\bar{P}+1)\ov 2}+
{\bar{l}(\bar{l}-2)\ov
4\nu^{2}})|\bar{P},\bar{l}>.
\label{1.6KMbar}
\enr

 The states with nonzero picture numbers can be described
explicitly using the following representation of the fields
$a,a^{+}$ ~\cite{FMS}. Let $\al(z),\bt(z)$ be the scalar free fields
with the OPE's
\be
\bt(z_{1})\bt(z_{2})=-\al(z_{1})\al(z_{2})=\ln z_{12}+r.
\label{1.bos1}
\en
Then
\ber
a=\exp(\al-\bt), \ a^{+}=\exp(-\al)\d\exp(\bt),
\label{1.bos2}
\enr
and the state $|P,l>$ corresponds to the vertex operator
\be
A_{P}(z)V_{l}(z)=\exp(P\al(z))\exp(-\im {l\ov \sqrt{2}\nu}\phi(z)).
\label{1.bos3}
\en
A similar formula is valid for the state $|\bar{P},\bar{l}>$.

 Let $W_{P,l}$ be the Fock module
generated from $|P,l>$ by the creation operators.
Analogously, we denote by $\bar{W}_{\bar{P},\bar{l}}$ the Fock
module generated from  $|\bar{P},\bar{l}>$ by the creation
operators of the right-moving fields $\bar{a},\bar{a}^{+},
\bar{\phi}$.

 In the tensor product $W_{P,l}\otimes\bar{W}_{\bar{P},\bar{l}}$ we
are going to construct Ishibashi state $|P,\bar{P},l,\bar{l}>>$
fulfilling on the boundary $z\bar{z}=1$ the relations ($g$-Ward identities)
\ber
(dzE(z)-d\bar{z}\bar{E}(\bar{z}))|P,\bar{P},l,\bar{l}>>=0, \nmb
(dzH(z)-d\bar{z}\bar{H}(\bar{z}))|P,\bar{P},l,\bar{l}>>=0, \nmb
(dzF(z)-d\bar{z}\bar{F}(\bar{z}))|P,\bar{P},l,\bar{l}>>=0.
\label{1.7}
\enr
Thus, boundary CFT under consideration lives on the unit disc
in $z$-plane.

 It follows from the Sugawara formula that
boundary conditions (\ref{1.7}) are conformally invariant
\ber
((dz)^{2}T-(d\bar{z})^{2}\bar{T})|P,\bar{P},l,\bar{l}>>=0, or \nmb
(L(n)-\bar{L}(-n))|P,\bar{P},l,\bar{l}>>=0.
\label{1.Vir}
\enr

 Let us solve (\ref{1.7}) in terms of the free fields.
Using (\ref{1.2d}), the similar formulas for the right-moving
sector the third relation from (\ref{1.7})
can be written as
\ber
(a^{+}(z)+\bar{a}^{+}(z))|P,\bar{P},l,\bar{l}>>=0.
\label{1.8C}
\enr
The second equation from (\ref{1.7}) impose a relation between
$a,\bar{a},\phi,\bar{\phi}$. The equations
which are consistent with (\ref{1.4}), (\ref{1.8C}) are given by
\ber
(a(z)-\bar{a}(\bar{z}))|P,\bar{P},l,\bar{l}>>=0, \nmb
(dz(\d \phi-\im {\sqrt{2}\ov \nu}(P+\bar{P}) z^{-1})+
d\bar{z}\bar{\d}\bar{\phi})|P,\bar{P},l,\bar{l}>>=0,
\label{1.9C}
\enr
The relations (\ref{1.9C}) can be rewritten in terms of the modes:
\ber
(a^{+}(n)+\bar{a}^{+}(-n))|P,\bar{P},l,\bar{l}>>=0,\nmb
(a(n)-\bar{a}(-n))|P,\bar{P},l,\bar{l}>>=0, \nmb
(\phi(n)-\bar{\phi}(-n)-
\im {\sqrt{2}\ov \nu}(P+\bar{P}) \dlt _{n,0})|P,\bar{P},l,\bar{l}>>=0.
\label{1.10C}
\enr

 The solution of (\ref{1.10C}) is given by
\ber
\bar{P}=-1-P, \ \bar{l}=l+2, \nmb
|P,-1-P,l,l+2>>=\exp(\Sgm_{m=-P}a(-m)\bar{a}^{+}(-m)+
\Sgm_{m=1+P}a^{+}(-m)\bar{a}(-m)) \nmb
\exp(\Sgm_{m=1}{1\ov m}\phi(-m)\bar{\phi}(-m))|P,-1-P,l,l+2>.\nmb
\label{1.11C}
\enr
and it fulfills the equations (\ref{1.7}).

 One needs also to make a consistency check of the relations
(\ref{1.10C}) and (\ref{1.Vir}). Due to the commutation
relations
\ber
\lbr L(n),a^{+}(m)\rbr=-ma^{+}(n+m), \nmb
\lbr L(n),a(m)\rbr=-(n+m)a(n+m), \nmb
\lbr L(n),\phi(m)\rbr=+\im {m(m-1)\ov \sqrt{2}\nu}\dlt_{n+m,0}-
m\phi(n+m), \nmb
\lbr \bar{L}(n),\bar{\phi}(m)\rbr=
-\im {m(m-1)\ov \sqrt{2}\nu}\dlt_{n+m,0}-
m\bar{\phi}(n+m)
\label{1.ch1C}
\enr
we see that (\ref{1.Vir}) is consistent with the relations
(\ref{1.10C}) for the fields $a,a^{+}$. For the field
$\phi$ we have
\ber
\lbr L(n)-\bar{L}(-n),\phi(m)-\bar{\phi}(-m)+
\im\sqrt{2}\nu \et\dlt_{m,0}\rbr|P,\bar{P},l,\bar{l}>>=\nmb
-m(\phi(n+m)-\bar{\phi}(-n-m)+
\im {\sqrt{2}\ov \nu}\dlt_{n+m,0})|P,\bar{P},l,\bar{l}>>=0
\label{1.ch2C}
\enr
which is consistent with (\ref{1.10C}) and (\ref{1.11C}).

 The boundary state (\ref{1.11C}) is a coherent state
in the tensor product $W_{0,l}\otimes \bar{W}_{-1,l+2}$.
In what follows we shall omit the right-moving indexes
in the notation of Ishibashi states (\ref{1.11C}) because
they are determined by the left-moving ones.

 For each pair of Ishibashi states $|P,l>>$, $|P',l'>>$ one can
calculate the cylinder partition function
\be
Z_{(P',l'),(P,l)}(\tau,\tt)=
<<P',l'|q^{(L(0)-c/24)}u^{H(0)}|P,l>>,
\label{1.13C}
\en
where $q=\exp(\im 2\pi \tau)$, $u=\exp(\im \pi \tt)$,
$\tau$ and $\tt$ are the complex numbers ($Im\tau >0$).
After the calculation of  the contractions we obtain
\ber
Z_{(P',l'),(P,l)}=\dlt_{(P,l),(P',l')}\nmb
q^{\Dl(P,l)-c/24}u^{2P-l}
\Pi_{m=-P}{1\ov (1-q^{m}u^{2})}\Pi_{m=P+1}{1\ov (1-q^{m}u^{-2})}
\Pi_{m=1}{1\ov (1-q^{m})},
\label{1.16}
\enr
where $\Dl (P,l)=-{P(P+1)\ov 2}+{l(l+2)\ov 4\nu^{2}}$.
It can be rewritten as the character $Ch(W_{0,l})$
of the Wakimoto module $W_{0,l}$
\ber
Z_{(P',l'),(P,l)}=(-1)^{P}\dlt_{(P,l),(P',l')}\nmb
q^{l(l+2)\ov 4\nu^{2}}u^{-l}
\Pi_{m=0}{1\ov (1-q^{m}u^{2})}\Pi_{m=1}{1\ov (1-q^{m}u^{-2})}
{1\ov (1-q^{m})}=\nmb
(-1)^{P}\dlt_{(P,l),(P',l')}Ch(W_{0,l}).
\label{1.17}
\enr

\vskip 10pt
\centerline{\bf3. Ishibashi states in irreducible
$\hat {su}(2)$-modules}
\centerline{\bf and boundary states in $SU(2)$ WZW model}
\vskip 10pt
 In this section we represent the free field construction of
Ishibashi and boundary states for irreducible integrable
$\hat{su}(2)$ representations. Thus we restrict the level $k$
and the momentum $l$ to be nonnegative integer numbers
such that $l<k+1$.

 To begin with the known BRST construction
of the irreducible $\hat{su}(2)$
representations.

 For each module $W_{0,l}$ one can associate
the complex (Felder's resolution)
~\cite{BFeld},~\cite{FFr},~\cite{BMP}
\ber
\cdots \rightarrow C^{{\infty\ov 2}-2}_{l}
\rightarrow C^{{\infty\ov 2}-1}_{l}
\rightarrow C^{{\infty\ov 2}}_{l}\rightarrow
C^{{\infty\ov 2}+1}_{l}\rightarrow C^{{\infty\ov 2}+2}_{l}
\rightarrow \cdots,\nmb
C^{{\infty\ov 2}+2n}_{l}=W_{0,l_{2n}},\ l_{2n}=l-2n\nu^{2}, \nmb
C^{{\infty\ov 2}+2n+1}_{l}=W_{0,l_{2n+1}},\ l_{2n+1}=-l-2-2n\nu^{2},
\ n\in Z,
\label{2.1}
\enr
with the differentials (BRST operators)
\ber
d_{2n}=Q_{l+1}, \ d_{2n+1}=Q_{\nu^{2}-l-1}, \nmb
Q_{m}=
\kp_{m} \oint_{\Gm} \prod_{i=1}^{m}dz_{i}S(z_{i}), \
\kp_{m}={1\ov m}
{\exp(\im 2\pi{m\ov \nu^{2}})-1\ov\exp(\im 2\pi{1\ov \nu^{2}})-1},
\label{2.2}
\enr
where
\ber
S(z)=a^{+}(z)\exp(\im {\sqrt{2} \ov \nu}\phi)(z)
\label{2.3}
\enr
and the integration contours $\Gamma$ are chosen similar to
~\cite{BFeld}: the integrations contours over the variables
$z_{2},...,z_{m}$ form a set nonintersecting curves going
counterclockwise from $z_{1}$ to $\exp(\im 2\pi)z_{1}$
and are localized in the neighborhood of the circle of radius
$|z_{1}|$ centered at the origin, which is the $z_{1}$ integration
contour.

 The irreducible $\hat{su}(2)$-module $M_{l}$ is given by the
cohomology groups of the complex (\ref{2.1})
~\cite{BFeld},~\cite{FFr}
\ber
H^{{\infty\ov 2}+i}=\dlt_{i,0}M_{l}.
\label{2.1h}
\enr
Thus the left-moving part of the Hilbert space of $SU(2)$ WZW
model is given by BRST invariant states (modulo BRST exact).
A similar statement takes place for the right-moving sector.

 Due to this result one can write the character $Ch_{l}(q,u)$
of the module $M_{l}$ as the Euler characteristic of the
complex (\ref{2.1}) ~\cite{BFeld}
\ber
Ch_{l}(q,u)=\Sgm_{k\in
Z}(Ch(W_{0,l-2k\nu^{2}})-
Ch(W_{0,-l-2-2k\nu^{2}}))(q,u).
\label{2.1ch}
\enr

 Hence, it is natural to suggest the free-field
construction of the Ishibashi state $|M_{l}>>$ for the module
$M_{l}$ as a superposition of the states (\ref{1.11C}) with $P=0$
\ber
|M_{l}>>=\Sgm_{k\in Z}
(c_{2k}|0,l-2k\nu^{2}>>+c_{2k+1}|0,-l-2-2k\nu^{2}>>),
\label{2.4a}
\enr
The coefficients $c_{n}$ can be fixed partly from the condition
\ber
<<M_{l_{1}}|(-1)^{f}q^{(L(0)-c/24)}u^{H(0)}|M_{l_{2}}>>=
\dlt_{l_{1},l_{2}}Ch_{l_{1}}(q,u),
\label{2.5}
\enr
where $f$ is the "ghost" number operator associated with
the complex (\ref{2.1}) and determined by
\be
f|v_{n}>=n|v_{n}>,\ |v_{n}>\in C^{{\infty\ov 2}+n}_{l}.
\label{2.gh}
\en

 Indeed, using (\ref{1.17}) we obtain
\ber
<<M_{l_{1}}|(-1)^{f}q^{(L(0)-c/24)}u^{H(0)}|M_{l_{2}}>>=\nmb
\Sgm_{k\in Z}(c^{2}_{2k}Ch(W_{0,l_{1}-2k\nu^{2}})-
c^{2}_{2k+1}Ch(W_{0,-l_{1}-2-2k\nu^{2}}))(q,u).
\label{2.6}
\enr
Comparing with (\ref{2.1ch}) we obtain
\be
c_{n}=\pm 1,\ n\in Z.
\label{2.4b}
\en
Thus the state (\ref{2.4a}) is a good candidate for the
free field realization of the Ishibashi
state in $SU(2)$ WZW model ~\cite{IW}. It would be a genuine
Ishibashi state
for the module $M_{l}$ if it did not radiate nonphysical closed
string states which are present in the free field representation
of the model. In other words, the overlap of this state with an
arbitrary closed string state which does not belong to
the Hilbert space of the WZW model should vanish. As we will see
this condition can be formulated as a BRST invariance condition
of the state (\ref{2.4a}) and it will fix the coefficients
$c_{n}$ up to the common sign.

 To investigate BRST invariance of the state
(\ref{2.4a}) one needs to
consider the structure of the Wakimoto modules.
We start with the module $W_{0,l}$.
Its vacuum vector $|0,l>$ has the properties
\ber
F(0)|0,l>=0, \nmb
H(0)|0,l>=-l|0,l>, \nmb
E(0)|0,l>=la(0)|0,l>, \nmb
(E(0))^{2}|0,l>=l(l-1)(a(0))^{2}|0,l>,
\label{1.12C}
\enr
where we have used the relation
\be
\lbr E(n),a(m)\rbr=-(a)^{2}(n+m).
\en
Thus we see that
\ber
(a(0))^{l+1}|0,l> is\ a\ cosingular\ vector, \
(E(0))^{l+1}|0,l>=0.
\label{1.coC}
\enr
We have also
\ber
(F(-1))^{\nu^{2}-l-1}|0,l>=\nmb
(a^{+}(-1))^{\nu^{2}-l-1}|0,l> is\ a\ singular\ vector.
\label{1.siC}
\enr
The whole structure of singular and cosingular vectors can be read
of from the determinant formula for the module $W_{0,l}$ ~\cite{Fr}
\ber
D_{\mu}=\Pi_{m,n>0}[-l+m\nu^{2}-(n+1)]^{p(\mu-n(m\dlt-\al))},\nmb
C_{\mu}=\Pi_{m,n>0}[l+(m-1)\nu^{2}-(n-1)]^{p(\mu-n((m-1)\dlt+\al))},
\label{1.detC}
\enr
where $\mu=r\dlt+s\al, r=0,1,...,s\in Z$ is a positive root of
$\hat{sl}(2)$, $\dlt$ is the imaginary root, $\al$ is the
positive root of $sl(2)$, $p(\mu)$ is the number of partitions
of $\mu$ into the sum of positive roots of $\hat{sl}(2)$.
The zeroes of $C_{\mu}$ correspond to the cosingular vectors,
while the zeroes of $D_{\mu}$ correspond to the singular vectors.
The complex (\ref{2.1}) gives the structure of submodules
of $W_{0,l}$.

 Let us consider the structure of $\bar{W}_{-1,\bar{l}}$,
$\bar{l}=l+2$.
Its vacuum vector $|-1,\bar{l}>$ has the properties
\ber
\bar{E}(0)|-1,\bar{l}>=0, \nmb
\bar{H}(0)|-1,\bar{l}>=(-2+\bar{l})|-1,\bar{l}>=l|-1,\bar{l}>, \nmb
\bar{E}(-1)|-1,\bar{l}>=
(k+2-\bar{l})\bar{a}(-1)|-1,\bar{l}>, \nmb
(\bar{E}(-1))^{2}|-1,\bar{l}>=
(k+2-\bar{l})(k+2-\bar{l}-1)|-1,\bar{l}>.
\label{1.12barC}
\enr
So we find that
\ber
(\bar{a}(-1))^{\nu^{2}-\bar{l}+1}|-1,\bar{l}>=
(\bar{a}(-1))^{\nu^{2}-l-1}|-1,\bar{l}>
is\ a\ cosingular\ vector, \nmb
(\bar{E}(-1))^{\nu^{2}-l-1}|-1,\bar{l}>=0.
\label{1.barcoC}
\enr
We have also
\ber
(\bar{F}(0))^{l+1}|-1,\bar{l}>= \nmb
(\bar{a}^{+}(0))^{l+1}|-1,\bar{l}> is\ a\ singular\ vector.
\label{1.barsiC}
\enr
The determinant formula for
the module $\bar{W}_{-1,\bar{l}}$ is given by ~\cite{Fr}
\ber
C_{\mu}=\Pi_{m,n>0}[-l+m\nu^{2}-(n+1)]^{p(\mu-n(m\dlt+\al))},\nmb
D_{\mu}=\Pi_{m,n>0}[l+(m-1)\nu^{2}-(n-1)]^{p(\mu-n((m-1)\dlt-\al))}.
\label{1.bardetC}
\enr

 Hence, we see that $\bar{W}_{-1,l+2}$ is dual to
$W_{0,l}$ ~\cite{Fr} and its Felder's complex can be written
as follows
\ber
\cdots \leftarrow \bar{C}^{{\infty\ov 2}+2}_{l}
\leftarrow \bar{C}^{{\infty\ov 2}+1}_{l}
\leftarrow \bar{C}^{{\infty\ov 2}}_{l}\leftarrow
\bar{C}^{{\infty\ov 2}-1}_{l}\leftarrow \bar{C}^{{\infty\ov 2}-2}_{l}
\leftarrow \cdots,\nmb
\bar{C}^{{\infty\ov 2}+2n}_{l}=
\bar{W}_{-1,\bar{l}_{2n}},\ \bar{l}_{2n}=2+l+2n\nu^{2}, \nmb
\bar{C}^{{\infty\ov 2}+2n-1}_{l}=
\bar{W}_{-1,\bar{l}_{2n-1}},\ \bar{l}_{2n-1}=-l+2n\nu^{2},
\ n\in Z,
\label{2.7}
\enr
where the differentials are given by
\ber
\bar{d}_{2n-1}=\bar{Q}_{l+1}, \
\bar{d}_{2n}=\bar{Q}_{\nu^{2}-l-1}, \nmb
\bar{Q}_{m}=
\kp_{m}
\oint_{\Gm^{-1}} \prod_{i=1}^{m}d\bar{z}_{i}\bar{S}(\bar{z}_{i}),\
\label{2.8}
\enr
where $n\in Z$,
\ber
\bar{S}(\bar{z})=\bar{a}^{+}(\bar{z})
\exp(-\im {\sqrt{2} \ov \nu}\bar{\phi})(\bar{z}),
\label{2.9}
\enr
and the integration contours $\Gm^{-1}$ have opposite
orientation and opposite ordering with respect to $\Gm$.

 Next we form a tensor product of the
complexes (\ref{2.1}) and (\ref{2.7}):
\ber
\cdots \rightarrow A^{{\infty\ov 2}-2}_{l}
\rightarrow A^{{\infty\ov 2}-1}_{l}
\rightarrow A^{{\infty\ov 2}}_{l}\rightarrow
A^{{\infty\ov 2}+1}_{l}\rightarrow A^{{\infty\ov 2}+2}_{l}
\rightarrow \cdots,\nmb
A^{{\infty\ov 2}+p}_{l}=
\bigoplus_{n+m=p}
C^{{\infty\ov 2}+n}_{l}\otimes \bar{C}^{{\infty\ov 2}+m}_{l}
\label{2.23}
\enr
with the differential $D$ defined by
\be
D_{p}(v_{n}\otimes \bar{v}_{m})=d_{n}v_{n}\otimes \bar{v}_{m}+
(-1)^{n}v_{n}\otimes \bar{d}_{m}\bar{v}_{m},\ n+m=p
\label{2.24}
\en
where $v_{n}\otimes \bar{v}_{m}$ is an arbitrary element
from $A^{{\infty\ov 2}+p}_{l}$.

 The cohomology groups of the complex (\ref{2.23}) are given by
\ber
{\bf H}^{{\infty\ov 2}+i}=\dlt_{i,0}M_{l}\otimes M^{*}_{l},
\label{2.23h}
\enr
where $M^{*}_{l}$ is dual module to $M_{l}$.

 The Ishibashi state we are looking for
can be considered as a linear functional on the Hilbert space of
$SU(2)$ WZW model, then it has to be an element from homology
group ${\bf H}_{\infty\ov 2}$. Therefore, the BRST invariance
condition for the state can be formulated as follows.

 Let us define the action of the differential $D$ on the state
$|M_{l}>>$ by the formula
\be
<<D^{*}M_{l}|v_{n}\otimes \bar{v}_{p-n}>\equiv
<<M_{l}|D_{p}|v_{n}\otimes \bar{v}_{p-n}>,
\label{2.24*}
\en
where $v_{n}\otimes \bar{v}_{p-n}$ is an arbitrary element
from $A^{{\infty\ov 2}+p}_{l}$. Then, BRST invariance condition
means that
\be
D^{*}|M_{l}>>=0.
\label{2.25}
\en
\vskip 10pt
\leftline{\bf Proposition.}
The superposition (\ref{2.4a}) satisfy the BRST invariance condition
(\ref{2.25}) if the coefficients $c_{n}$
obey the following equation
\be
(-1)^{-n}c_{-n}+c_{1-n}=0.
\label{2.25a}
\en
Note that the last relation is consistent with (\ref{2.4b})
and we obtain two sets of solutions of (\ref{2.25a})
\ber
c_{2m}=(-1)^{m},\ c_{2m+1}=-(-1)^{m},\ or \nmb
c_{2m}=-(-1)^{m},\ c_{2m+1}=(-1)^{m},\ m\in Z.
\label{2.25b}
\enr

 Now we move on to the prove of the Proposition.

 Let us consider an arbitrary state
\be
|w_{-n}\otimes\bar{w}_{-m}>\in A^{{\infty\ov 2}-n-m}_{l}
\label{2.28}
\en
from the complex (\ref{2.23}). In view of (\ref{2.4a}) only the
states from $A^{\infty\ov 2}_{l}$ have nonzero overlap with
Ishibashi state $|M_{l}>>$. Because of the differential
$D$ rises the ghost number by one we have to put in (\ref{2.28})
$-n-m=-1$. So one needs to show that
\ber
<<M_{l}|D_{-1}|w_{-n}\otimes \bar{w}_{-1+n}>\equiv \nmb
<<M_{l}|d_{-n}+(-1)^{-n}\bar{d}_{-1+n}|w_{-n}\otimes \bar{w}_{-1+n}>=0.
\label{2.29}
\enr

 We will imply in the following that the state (\ref{2.28})
 corresponds to the field $(D_{-1}(w_{-n}\otimes \bar{w}_{-1+n}))(z,\bar{z})$
 which is placed at the center $z=\bar{z}=0$ of the disk.
 
 Let us consider the first term  of (\ref{2.29}).
\ber
<<M_{l}|d_{-n}|w_{-n}\otimes \bar{w}_{-1+n}>=
\kp_{-n}<<M_{l}|\oint_{\Gm}\prod_{i=1}^{N}dz_{i}S(z_{i})
|w_{-n}\otimes \bar{w}_{-1+n}>=\nmb
\kp_{-n}<<M_{l}|\oint_{\gm_{1}}dz_{1}S(z_{1})
\int_{\Gm'}\prod_{i=2}^{N}dz_{i}S(z_{i})
|w_{-n}\otimes \bar{w}_{-1+n}>.
\label{2.30}
\enr
In this formula $N$ is determined by (\ref{2.2}). Recall also that
we can chose $z_{1}$ integration contour
$\gm_{1}$ as the unit circle centered at $z=\bar{z}=0$
(so it coincide with the boundary of the disk) such that
the integration contours $\Gm'$ form a set of non-intersecting
nested curves going counterclockwise from $z_{1}$ to
$\exp(\im 2\pi)z_{1}$ and are localized in the neighborhood
of the contour $\gm_{1}$. When acting on
$|w_{-n}\otimes \bar{w}_{-1+n}>$ the BRST current
\be
J(z_{1})=S(z_{1})\int_{\Gm'}\prod_{i=2}^{N}dz_{i}S(z_{i})
\label{2.brst}
\en
is single valued around the center of the disk.

 To prove (\ref{2.29}) we change first $dz_{1}S(z_{1})$
taking into account normal ordering in the
$\exp(\im {\sqrt{2}\ov \nu}\phi)$
and using the relations (\ref{1.10C})
\ber
<<M_{l}|dz_{1}S(z_{1})=\nmb
<<M_{l}|dz_{1}a^{+}(z_{1})
\exp(\im {\sqrt{2}\ov\nu}\phi_{<}(z_{1}))
\exp(\im {\sqrt{2}\ov\nu}\phi_{0})
z_{1}^{\im {\sqrt{2}\ov\nu}\phi(0)}
\exp(\im {\sqrt{2}\ov\nu}\phi_{>}(z_{1}))=\nmb
<<M_{l}|\exp(\im {\sqrt{2}\ov\nu}(\phi_{0}+\bar{\phi}_{0}))
d\bar{z}_{1}\bar{S}(\bar{z}_{1}).
\label{2.30a}
\enr
Then, deform the first integration contour from $\Gm'$ towards
the boundary $z\bar{z}=1$ and apply
the relations (\ref{1.10C}). Note that we have
no relations for the modes $\phi_{0}$, $\bar{\phi}_{0}$
canonically conjugated to the momentum modes $\phi(0)$, $\bar{\phi}(0)$.
So we keep the $\phi_{0}$ unchanged during this process. Thus we
obtain
\ber
<<M_{l}|dz_{1}J(z_{1})|w_{-n}\otimes \bar{w}_{-1+n}>=\nmb
\kp_{-n}<<M_{l}|\exp(\im
{\sqrt{2}\ov\nu}2(\phi_{0}+\bar{\phi}_{0}))
d\bar{z}_{1}\bar{S}(\bar{z}_{1})
\oint_{\gm_{2}}d\bar{z}_{2}\bar{S}(\bar{z}_{2})
\int_{\Gm''}\prod_{i=3}^{N}dz_{i}S(z_{i})\nmb
|w_{-n}\otimes\bar{w}_{-1+m}>.
\label{2.31}
\enr

 Now we can deform the integration contour $\gm_{2}$ towards the
origin such that $|\bar{z}_{2}|<|z_{N}|$.

 Repeating the same procedure for other contours we obtain
\ber
<<M_{l}|d_{-n}|w_{-n}\otimes\bar{w}_{-1+n}>=\nmb
\kp_{-n}<<M_{l}|\exp(\im
{\sqrt{2}\ov\nu}N(\phi_{0}+\bar{\phi}_{0}))
\oint_{\Gm^{-1}}\prod_{i=N}^{1}d\bar{z}_{i}\bar{S}(\bar{z}_{i})
|w_{-n}\otimes\bar{w}_{-1+m}>=\nmb
c_{1-n}<<\bar{l}_{-1+n},l_{1-n},-1,0|
\exp(\im {\sqrt{2}\ov\nu}N(\phi_{0}+\bar{\phi}_{0}))
|w_{-n}\otimes\bar{d}_{-1+n}\bar{w}_{-1+n}>,
\label{2.32}
\enr
where $\Gm^{-1}$ denotes the set of nested contours with
opposite orientation and opposite ordering.

 The operator
$\exp(\im {\sqrt{2}\ov\nu}N(\phi_{0}+\bar{\phi}_{0}))$
only shifts the ghost numbers of the state
$|w_{-n}\otimes\bar{d}_{-1+n}\bar{w}_{-1+n}>$,
such that the only nonzero pairing appears for
$<<\bar{l}_{-1+n},l_{1-n},-1,0|$.
From the other hand
\ber
<<M_{l}|\bar{d}_{-1+n}|w_{-n}\otimes\bar{w}_{-1+n}>=\nmb
c_{-n}<<\bar{l}_{n},l_{-n},-1,0|
w_{-n}\otimes\bar{d}_{-1+n}\bar{w}_{-1+n}>.
\label{2.33}
\enr
Comparing these formulas we conclude that (\ref{2.29})
will be satisfied
iff (\ref{2.25a}) is fulfilled. It proves the Proposition.

 To complete the free field construction of the Ishibashi state
one needs to check that $|M_{l}>>$
is not BRST exact state. It can be done by projecting the
state $|M_{l}>>$ onto an element from the cohomology
class ${\bf H}^{{\infty\ov 2}}$.
Let us consider  the case when $l>0$ and choose the representative from
the zero-level states
of $M_{l'}\otimes M^{*}_{l'}$. It can be written as
$(E(0))^{n}(\bar{F}(0))^{m}|0,-1,l',l'+2>$. Then
\ber
<<M_{l}|(E(0))^{n}(\bar{F}(0))^{m}|0,-1,l',l'+2>=\nmb
(-1)^{n}l(l-1)...(l-n+1)\dlt_{n,m}\dlt_{l,l'}c_{0}.
\label{2.trace}
\enr
For the case $l=0$ we choose
$(F(-1))^{n}(\bar{E}(-1))^{m}|0,-1,l',l'+2>, n<k-l'+1$ as a representative
from $M_{l'}\otimes M^{*}_{l'}$. Projecting $|M_{0}>>$ onto the
representative we find the expression like (\ref{2.trace}).
Therefore, we see that result is not zero and hence the Ishibashi
state $|M_{l}>>$ defines a homology class. Moreover, similar
to ~\cite{FFFS} the
projection is given by the calculation of diagonal matrix
elements in the irreducible $su(2)$- representation if we put
\be
c_{0}=1.
\label{2.norm}
\en

In view of the Proposition we can obtain free field
representation
of the boundary states in $SU(2)$-WZW model just applying
the formula found by Cardy ~\cite{C}:
\ber
|B_{l}>>=\sum_{j\in I}{S_{lj}\ov \sqrt{S_{0j}}}|M_{j}>>,
\label{2.34}
\enr
where $S_{lj}, j,l\in I$ is the matrix of modular
transformation of $\hat {su}(2)$-characters:
\ber
S_{lj}={\sqrt{2}\ov \nu}\sin(\pi (j+1)(l+1)/\nu^{2}).
\label{2.35}
\enr

 Using (\ref{2.34}), (\ref{2.trace}) and (\ref{2.norm}) it is easy
to recover the wave function of the $SU(2)$ boundary state found in
~\cite{FFFS}.

\vskip 10pt
\centerline{\bf4. Free field realization of twisted boundary states.}
\vskip 10pt

 In this section we consider briefly the twisted boundary
conditions of automorphism type. The theory treating such
boundary conditions for arbitrary CFT's has been developed
in ~\cite{FuS} and some applications of these results to WZW models
has been considered in ~\cite{BFuS}. The automorphisms which are 
interested to us
here are induced by automorphisms of the algebra
$su(2)$. They are all inner automorphisms and can be
represented by the adjoint action of the group $SU(2)$. We
consider here only the case when it is given by
the Weyl reflection $r$ of $su(2)$. The generalization
is straightforward.

 Hence, we have to construct
first the twisted Ishibashi states $|0,-1,l,\bar{l}>>^{r}$
fulfilling on the
boundary $z\bar{z}=1$ the relations
\ber
(dz(r.E)(z)-d\bar{z}\bar{E}(\bar{z}))|0,-1,l,\bar{l}>>^{r}=0, \nmb
(dz(r.H)(z)-d\bar{z}\bar{H}(\bar{z}))|0,-1,l,\bar{l}>>^{r}=0, \nmb
(dz(r.F)(z)-d\bar{z}\bar{F}(\bar{z}))|0,-1,l,\bar{l}>>^{r}=0,
\label{4.1}
\enr
where
\be
(r.E)(z)=-F(z),\
(r.H)(z)=-H(z),\
(r.F)(z)=-E(z).
\label{4.2}
\en
The relations (\ref{4.1}) are equivalent to
\ber
(\bar{a}^{+}(z)+a^{2}a^{+}(z)+k\d a(z)-
\im \sqrt{2}\nu a\d \phi(z))|0,-1,l,\bar{l}>>^{r}=0,\nmb
(\bar{a}(z)+a^{-1}(z))|0,-1,l,\bar{l}>>^{r}=0,\nmb
(\d \bar{\phi}(z)-\d \phi(z)-
\im \sqrt{2}\nu (a^{-1}\d a(z)+{1\ov \nu^{2}}z^{-1}))|0,-1,l,\bar{l}>>^{r}=0,
\label{4.solr}
\enr
where $a^{-1}(z)\equiv\exp(-\al+\bt)(z)$.
In terms of the modes these equations are given by
\ber
(\bar{a}^{+}(n)-E(-n))|0,-1,l,\bar{l}>>^{r}=0,\nmb
(\bar{a}(n)+a^{-1}(-n))|0,-1,l,\bar{l}>>^{r}=0,\nmb
(\bar{\phi}(n)-\phi(-n)-
\im \sqrt{2}\nu ((a^{-1}\d a)(-n)+
\dlt_{-n,0}{1\ov \nu^{2}}))|0,-1,l,\bar{l}>>^{r}=0.
\label{4.solrn}
\enr
One can easily check that
\ber
\exp(E(0))\exp(-F(0))\exp(E(0))a(z)=-a^{-1}(z),\nmb
\exp(E(0))\exp(-F(0))\exp(E(0))a^{+}(z)=-F(z),\nmb
\exp(E(0))\exp(-F(0))\exp(E(0))\d \phi(z)=
\d \phi(z)+\im \sqrt{2}\nu a^{-1}\d a(z).
\label{4.formul}
\enr
Thus the Ishibashi state we are going to construct
has to act on the fields $\bar{a},\bar{a}^{+},\bar{\phi}$
as the composition
\be
(\bar{a},\bar{a}^{+},\bar{\phi}) \rightarrow
(a,a^{+},\phi) \rightarrow r(a,a^{+},\phi),
\label{4.3}
\en
where
\be
r=\exp(E(0))\exp(-F(0))\exp(E(0)).
\label{4.4}
\en
Then, in according to (\ref{4.3})
we can write
\be
|0,-1,l,\bar{l}>>^{r}=(r\otimes 1)|0,l>>.
\label{4.5}
\en
It is obvious that we can use instead of $r$ defined by
(\ref{4.4}) the operator $\bar{r}$ which acts in the right-moving
sector and defined by
\be
\bar{r}=\exp(\bar{F}(0))\exp(-\bar{E}(0))\exp(\bar{F}(0)).
\label{4.4bar}
\en
Then one can rewrite (\ref{4.5}) in the equivalent form
\be
|0,-1,l,\bar{l}>>^{r}=(1\otimes \bar{r})|0,l>>.
\label{4.5bar}
\en
It is easy to see also that operators $r$ and $\bar{r}$ change
the pictures, the picture of the state $r\otimes 1|0,l>>$
is $(-1,-1)$ and the picture of the state
$1\otimes \bar{r}|0,l>>$ is $(0,0)$. It enables us to conclude
that
\ber
(r\otimes 1)|0,l>>\in W_{-1,-2-l}\otimes \bar{W}_{-1,l+2},\nmb
(1\otimes \bar{r})|0,l>>\in W_{0,l}\otimes \bar{W}_{0,-l}.
\label{4.6}
\enr

 The next step is to construct twisted Ishibshi states for the
irreducible $\hat{su}(2)$ modules. Because of the differentials of
the Felder's complexes are invariant with respect to the left-moving
and right-moving $\hat{su}(2)$ algebras one can easily extend
the construction of the Section 3 to the case of twisted
boundary conditions.

 Let us consider for example the representation (\ref{4.5bar})
for the twisted Ishibashi state in Wakimoto modules.
In this case the structures of modules
$W_{0,l}$ and $\bar{W}_{0,-l}$ coincide so that
complex in the right-moving sector coincides
with the complex (\ref{2.1}) in the left-moving sector.
Similar to (\ref{2.23}) we can
form the double complex which calculates the tensor
product of irreducible representations. Thus it is easy to
see that state
\be
(1\otimes \bar{r})|M_{l}>>,
\label{4.7}
\en
where $|M_{l}>>$ is given by (\ref{2.4a}), (\ref{2.25a}),
is BRST invariant and gives free field realization of twisted
Ishibashi state in $SU(2)$ WZW model. It is obvious also that
the state
\be
(1\otimes \bar{r})|B_{l}>>,
\label{4.8}
\en
where $|B_{l}>>$ is given by (\ref{2.34}) represents
twisted boundary state of the model.

\vskip 10pt
\centerline{\bf5. Discussion}

 In this note we have constructed $\hat{su}(2)$ and
BRST invariant Ishibashi states
in $SU(2)$ WZW models using free field realization
of the bulk WZW model. Each Ishibashi state of the model is given by
infinite superposition of Ishibashi states of Wakimoto modules,
forming Felder's complex for irreducible
$\hat{su}(2)$ module. It is shown that coefficients of the
superposition are fixed uniquely by the BRST invariance condition
and the Ishibashi state constructed this way is not BRST exact
and hence represents nontrivial homolgy class.
We group these free field realized Ishibashi
states into the boundary states of $SU(2)$ WZW model using the
solution
found by Cardy. Due to BRST invariance the boundary states
do not radiate non-physical closed string
states (which are present originally in the free field realized
WZW model) and its wave functions coincide with the wave functions
of conjugacy classes found in ~\cite{FFFS}.

 Using free field
representation of the automorphism group
of $su(2)$ algebra we have constructed also
the twisted automorphism type boundary states
when the automorphism is given by Weyl reflection.

 Note also that free field representation for
the character of the irreducible $\hat{su}(2)$ module (\ref{2.6})
 
very close to the open-string Witten index
 ~\cite{DougF},~\cite{Qint}
which has been widely discussed in the literature in context
of $D$-branes on Calabi-Yau manifolds.
Indeed, using (\ref{2.6}, \ref{2.34}) and Verlinde formula
one can obtain:
\ber
<<B_{l_{1}}|(-1)^{f}q^{(L(0)-c/24)}u^{H(0)}|B_{l_{2}}>>=\nmb
\sum_{j\in I}{S_{l_{1}j}\ov S_{0j}}{S_{l_{2}j}\ov S_{0j}}
S_{0j}Ch_{j}(q,u)=
\sum_{j,l\in I}N_{l_{1}l_{2}}^{l}S_{lj}Ch_{j}(q,u)=\nmb
\sum_{l\in I}N_{l_{1}l_{2}}^{l}Ch_{l}(Sq,Su).
\label{5.1}
\enr

 Thus in the open string sector we have
\ber
Tr_{\Omega_{l_{1}l_{2}}}(-1)^{f^{op}}\tld{q}^{(L^{op}(0)-c/24)}
\tld{u}^{H^{op}(0)}=
\sum_{l\in I}N_{l_{1}l_{2}}^{l}Ch_{l}(\tld{q},\tld{u}),
\label{5.2}
\enr
where $\Omega_{l_{1}l_{2}}$ is Hilbert space in the open string
sector, $f^{op}$, $L^{op}(0)$, $H^{op}(0)$, are the corresponding
operators, and $\tld{q}=\exp(-\im {2\pi\ov\tau})$,
$\tld{u}=\exp(\im {\pi\theta\ov \tau})$.

 The right hand
side of this equality is given by the character-valued index
of the BRST operator in the space $\Omega_{l_{1},l_{2}}$.
This expression is similar to that obtained for example in
~\cite{Qint}, ~\cite{LW}, ~\cite{HIV} for
$N$=2 superconformal field theories.
It would be interesting to find geometric
interpretation of (\ref{5.2}).

 We close with a brief discussion of some
directions to develop.  The first one is
a generalization of BRST construction of Ishibashi states
for higher rank groups. It would also be interesting to join
BRST construction of Ishibashi states in WZW models and
quantum Drinfeld-Sokolov reduction ~\cite{DS} to
give free field
realization of Ishibashi and boundary states in the CFT's with $W$-algebra of
symmetries ~\cite{FL}.
The next obviously important direction is an extension of
the construction to the case of supersymmetric WZW models
and $N$=2 coset models.

\vskip 10pt
\centerline{\bf Acknowledgments}

 This work was supported in part by grants RBRF-01-0216686,
RBRF-96-1596821, INTAS-OPEN-97-1312, INTAS-00-0005 and RPI-2254.

\end{document}